# Encapsulation theory: the configuration efficiency limit.

Edmund Kirwan[*]
www.EdmundKirwan.com

**Abstract**

*This paper shows how maximum possible configuration efficiency of an indefinitely large software system is constrained by chosing a fixed upper limit to the number of program units per subsystem. It is then shown how the configuration efficiency of an indefinitely large software system depends on the ratio of the total number of informaiton hiding violational software units divided by the total number of program units.*

**1. Introduction**

The configuration efficiency of a software system was introduced in [1] and defined as being that proportion of potential structural complexity (P.S.C.) that a system expresses over and above the minimum possible P.S.C. of that system.

Two questions in [1] concerning the configuration efficiency, however, remained unanswered.

The first relates to figure 24 in the referenced paper, reproduced here:

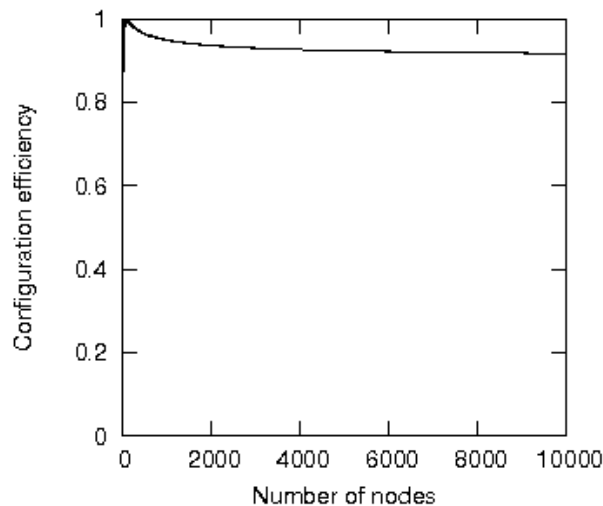

*Figure 24: Configuration efficiency with ten program units per subsystem*

What is the precise relationship between the configuration efficiency and the constraint of a fixed number of program units per subsystem, as show in figure above?

The second question relates to table 2 in the referenced paper, reproduced here:

---

|  | *Num. nodes* | *P.S.C.* | *C.E.* | *I.H.V.* |
|---|---|---|---|---|
| SEdit | 135 | 16956 | 0.09 | 93% |
| Fractality | 240 | 19042 | 0.86 | 25% |
| HSQLdb | 283 | 61633 | 0.37 | 72% |
| Jasper Reports | 316 | 99540 | 0 | 100% |
| JBPM | 366 | 133590 | 0 | 100% |
| Manta-Ray | 384 | 132298 | 0.12 | 89% |
| Blue Marine | 468 | 194991 | 0.13 | 89% |
| Jboss | 4244 | 17619836 | 0.02 | 98% |
| Eclipse | 39114 | 933916300 | 0.4 | 61% |

*Table 2: Sample systems and associated metrics*

Why, in table 2, is there a correlation between the configuration efficiency (C.E.) and the ratio of the number of public program units divided by the total number of program units (the I.H.V., expressed as a percentage)? (Recall that the `public` accessor in Java – in which all the systems above were written – is the means of implementing the more general notion of information hiding violation.)

This paper answers both of these questions.

Note that this paper covers only the non-hierarchical encapsulation context.

## 2. Fixed-sized subsystems

As software systems grow, the number of program units per subsystem must also grow if the system's P.S.C. is to be minimised and the configuration efficiency maximised. For example, a system of 10,000 program units, with one information hiding violation per subsystem, would require one hundred program units in each subsystem.

Modern software practices, however, tend to restrict the number of program units per subsystem to a figure much lower than 100. No figure is universally agreed upon, but subsystems typically contain fewer than fifty program units.

The question may then be asked: if a programmer restricts the number of program units per subsystem, how does this affect the P.S.C. expressed by the system? Is there a precise relationship between a fixed upper limit to the number of program units per subsystem and the maximum possible configuration efficiency of a system?

The answer to this last question is yes.

If a programmer restricts the number of program units per subsystem to $x$, say, and if $p$ is the regional information hiding violation (the number of public program units per subsystem) then the maximum possible configuration efficiency of that system as it grows indefinitely large is given by the equation (see theorem 2.3):

$$\lim_{n \to \infty} c_e = 1 - \frac{p}{x}$$

This maximum possible configuration efficiency of an indefinitely large system as is called the configuration efficiency limit.

Thus in the system discussed above, if the programmer restricts the number of program units per subsystem to fifty and with each subsystem containing one public program unit, then the configuration efficiency limit of that system is:

$$\lim_{n \to \infty} c_e = 1 - \frac{p}{x} = 1 - \frac{1}{50} = 0.98$$

If the programmer restricts the same system to 10 program units per subsystem, the configuration efficiency limit will be 0.9, as suggested by figure 24, shown in the introduction.

## 2. System information hiding.

Examing at the data in table 2, we see that Jboss, for example, has a configuration efficiency of 0.02 and the number of public classes expressed as a percentage of the total number of classes is 98%. Expressing this percentage as a ratio, we see that Jboss's public-to-total class ratio is 0.98, which seems to be one minus the configuration efficiency. Eclipse, similarly, as a configuration efficiency of 0.4 and a ratio of 0.61, again, approximately one minus the configuration efficiency.

The question may then be asked: is there a precise relationship between the maximum possible configuration efficiency of a system and the ratio of the number of information hiding violational program units divided by the total number of program units?

The answer is yes.

If the total number of a program units in a system is $n$ and the total number of information hiding violational program units is $|h(G)|$ then the configuration efficiency limit can also be expressed as (see theorem 2.4):

$$\lim_{n \to \infty} c_e = 1 - \frac{|h(G)|}{n}$$

The ratio in this equation is the ratio of the number of public classes divided by the total number of classes in a system. This equation only holds for large systems, which is why table data correlate more strongly for the larger than the smaller systems. This also presents a method for calculating a convenient approximation to the configuration efficiency for large systems.

### Conclusions

This paper suggests both that the modern practice of constraining packages sizes is a viable mechanism for managing the P.S.C. of indefinitely large systems, and that the maximum possible configuration efficiency of indefinitely large systems is related to the total number of information hiding violational program units.

### Appendix A

### *Theorem 2.1.*

Given a uniformly distributed, encapsulated graph $G$ of $n$ nodes and of $r$ encapsulated regions, with each encapsulated region having an information hiding violation of $p$, the configuration inefficiency $c_i$ is given by the equation:

$$c_i = \frac{\frac{n}{r} + rp - 2\sqrt{np}}{n - 2\sqrt{np} + p}$$

*Proof:*

From section 6.5.4. of [1], the configuration inefficiency is given by:

$$c_i = \frac{s(G) - s_{min}(G)}{s_{max}(G) - s_{min}(G)} \quad (i)$$

Where:

$s(G)$ = actual system P.S.C.

$s_{min}(G)$ = minimum system P.S.C.

$s_{max}(G)$ = maximum system P.S.C.

From theorem 1.1:

$$s_{max}(G) = n(n-1)$$

From theorem 1.8:

$$s(G) = n\left(\frac{n}{r} - 1 + (r-1)p\right)$$

From theorem 1.14:

$$s_{min}(G) = n(2\sqrt{np} - 1 - p)$$

Substituting all three equations into (i) gives:

$$c_i = \frac{n\left(\frac{n}{r} - 1 + (r-1)p\right) - n(2\sqrt{np} - 1 - p)}{n(n-1) - n(2\sqrt{np} - 1 - p)}$$

$$= \frac{\left(\frac{n}{r} - 1 + (r-1)p\right) - (2\sqrt{np} - 1 - p)}{(n-1) - (2\sqrt{np} - 1 - p)}$$

$$= \frac{\frac{n}{r} - 1 + rp - p - 2\sqrt{np} + 1 + p}{n - 1 - 2\sqrt{np} + 1 + p}$$

$$= \frac{\frac{n}{r} + rp - 2\sqrt{np}}{n - 2\sqrt{np} + p}$$

*QED*

## Theorem 2.2.

Given a uniformly distributed, encapsulated graph of *n* nodes and of *r* encapsulated regions, with each encapsulated region having an information hiding violation of *p,* the configuration efficiency $c_e$ is given by the equation:

$$c_e = \frac{n + p + \frac{n}{r} - rp}{n - 2\sqrt{np} + p} \quad \text{(i)}$$

*Proof:*

From section 6.5.4. of [1], the configuration efficiency is given by:

$$c_e = 1 - c_i$$

From theorem 2.1:

$$c_i = \frac{\frac{n}{r} + rp - 2\sqrt{np}}{n - 2\sqrt{np} + p}$$

Substituting into (i) gives:

$$c_e = 1 - \frac{\frac{n}{r} + rp - 2\sqrt{np}}{n - 2\sqrt{np} + p}$$

$$= \frac{n - 2\sqrt{np} + p}{n - 2\sqrt{np} + p} - \frac{\frac{n}{r} + rp - 2\sqrt{np}}{n - 2\sqrt{np} + p}$$

$$= \frac{n - 2\sqrt{np} + p - \frac{n}{r} - rp + 2\sqrt{np}}{n - 2\sqrt{np} + p}$$

$$= \frac{n + p - \frac{n}{r} - rp}{n - 2\sqrt{np} + p}$$

*QED*

## Theorem 2.3.

Given a uniformly distributed, encapsulated graph of *n* nodes and of *r* encapsulated regions, with each encapsulated region having an information hiding violation of *p,* and given that the $i^{th}$ encapsulated region $K_i$ contains the fixed number *x* nodes, the limit of the configuration efficiency as the graph grows indefinitely large is given by:

$$\lim_{n \to \infty} c_e = 1 - \frac{p}{x}$$

*Proof:*

From theorem 2.2:

$$c_e = \frac{n+p-\frac{n}{r}-rp}{n-2\sqrt{np}+p} \quad (i)$$

By definition:

$$x = \frac{n}{r} \text{ and so } r = \frac{n}{x}$$

Substituting both of these equations into (i) gives:

$$c_e = \frac{n+p-x-\frac{np}{x}}{n-2\sqrt{np}+p}$$

$$= \frac{\frac{n}{n}+\frac{p}{n}-\frac{x}{n}-\frac{np}{nx}}{\frac{n}{n}-2\sqrt{\frac{p}{n}}+\frac{p}{n}}$$

$$= \frac{1+\frac{p}{n}-\frac{x}{n}-\frac{p}{x}}{1-2\sqrt{\frac{p}{n}}+\frac{p}{n}}$$

Taking the limit as $n$ tends to infinity gives:

$$\lim_{n \to \infty} c_e = \frac{1-\frac{p}{x}}{1} = 1-\frac{p}{x}$$

*QED*

## Theorem 2.4

Given a uniformly distributed, encapsulated graph $G$ of $n$ nodes and of $r$ encapsulated regions, with each encapsulated region having an information hiding violation of $p$, and given that $G$ has an information hiding violation of $h(G)$, the limit of the configuration efficiency as the graph grows indefinitely large is given by:

$$\lim_{n \to \infty} c_e = 1 - \frac{|h(G)|}{n}$$

*Proof:*

From theorem 2.3, where the $i^{th}$ encapsulated region $K_i$ contains the fixed number $x$ nodes, the limit of the configuration efficiency of $G$ as the graph grows indefinitely large is given by:

$$\lim_{n \to \infty} c_e = 1 - \frac{p}{x} \quad \text{(i)}$$

By definition:

$$|K_i| = x = \frac{n}{r}$$

Substituting this into (i) gives:

$$\lim_{n \to \infty} c_e = 1 - \frac{rp}{n} \quad \text{(ii)}$$

But by item (iii) of definintion D22 in [1]:

$$p = |h(K_i)| = \frac{|h(G)|}{r}$$

Substituting this into (ii) gives:

$$\lim_{n \to \infty} c_e = 1 - \frac{r|h(G)|}{rn} = 1 - \frac{|h(G)|}{n}$$

*QED*